# Bright Room-Temperature Single Photon Emission from Defects in Gallium Nitride

*Amanuel M. Berhane[1], †, Kwang-Yong Jeong[2], †, Zoltán Bodrog[3], Saskia Fiedler[1], Tim Schröder[2], Noelia Vico Triviño[2], Tomás Palacios[2], Adam Gali[3], Milos Toth[1], Dirk Englund[2]\* and Igor Aharonovich[1]\**

† *These authors contributed equally.*

1. School of Mathematical and Physical Sciences, University of Technology Sydney, Ultimo, New South Wales 2007, Australia.

2. Department of Electrical Engineering and Computer Science, MIT, Cambridge, MA 02139, USA.

3. Institute for Solid State Physics and Optics, Wigner RCP of the Hungarian Academy of Sciences, Hungary.

E-mail: igor.aharonovich@uts.edu.au ; englund@mit.edu

ABSTRACT

Single photon emitters play a central role in many photonic quantum technologies[1, 2]. A promising class of single photon emitters consists of atomic color centers in wide-bandgap crystals, such as diamond[3, 4] silicon carbide[5] and hexagonal boron nitride[6]. However, it is currently not possible to grow these materials as sub-micron thick films on low-refractive index substrates, which is necessary for mature photonic integrated circuit technologies. Hence, there is great interest in identifying quantum emitters in technologically mature semiconductors that are compatible with suitable heteroepitaxies. Here, we demonstrate robust single photon emitters based on defects in gallium nitride (GaN), the most established and well understood semiconductor that can emit light over the entire visible spectrum. We show that the emitters have excellent photophysical properties including a brightness in excess of $500 \times 10^3$ counts/s. We further show that the emitters can be found in a variety of GaN wafers, thus offering reliable and scalable platform for further technological development. We propose a theoretical model to explain the origin of these emitters based on cubic inclusions in hexagonal gallium nitride. Our results constitute a feasible path to scalable, integrated on-chip quantum technologies based on GaN.

**MAIN TEXT**

III-nitrides are widely used in solid state lighting[7], high-frequency and high-power electronics[8, 9] and laser technologies[10]. In particular, gallium nitride (GaN) features advantageous optical and electronic properties such as non-linear second order susceptibility[11], spontaneous and piezoelectric polarization[12], biocompatibility[13] and a direct, wide bandgap[14, 15]. Thus, GaN is increasingly used as platform for photonic integrated circuits (PIC)[16, 17] including waveguides[18], microdisk cavities[19-21], and photonic crystals[22-24]. In addition, GaN is the basis of a multi-million dollar efficient lighting industry underpinned by mature nanofabrication and growth technologies[25, 26].

In this work, we report room temperature (RT), bright, stable single photon emitters (SPEs) in GaN films that do not require any post-growth sample treatments. The emitters are defects that are optically active in the visible/near-infrared (NIR) spectral range, and the zero-phonon lines (ZPL) span a wide range of wavelengths. They were found in five GaN wafers that have different doping types and levels, and are grown on various substrates using Metal Organic Chemical Vapor Deposition (MOCVD), the most common commercially viable technique for the growth of device-grade GaN. The pervasiveness of the SPEs in different GaN samples demonstrates that our finding are not limited to a rare defect found in select few ultrapure samples.

To identify and characterize the SPEs, we analyzed 5 GaN samples grown on sapphire and silicon carbide (SiC). Sample A consists of a 4 µm thick Magnesium (Mg)-doped GaN film grown on sapphire. For clarity, we present an optical image and an atomic force microscope (AFM) scan of Sample A in **Figure 1**(a-b), while a summary of the structural characterization of the other samples is provided in the supporting information (table S1). The crystalline quality of the GaN samples was evaluated by X-ray diffraction (XRD) measurements, confirming their single crystal nature (see table S2). For all samples, surface roughness is below 1 nm (see table S1, S2).

The spectroscopy measurements on all samples were performed using a scanning confocal microscope (microscope objective with numerical aperture 0.9) with green laser excitation (532nm) focused to a laser spot with a diameter of ~ 400 nm. **Figure 1**c shows a representative 60 µm X 60 µm fluorescence map with localized, bright spots labeled E1-E8, corresponding to isolated emission spots in Sample A. This map is obtained from a 1 $cm^2$ sample using an excitation power of 300 µW (measured before the microscope objective). An analysis of other regions of sample A indicate an average of 1 emitter per 25 $\mu m^2$.

**Figure 1**d summarizes fluorescence scans of GaN Samples B-E, which consist of the layer structures shown in the insets. Sample B was grown on sapphire, C and D on sapphire overgrown with different GaN/InGaN epilayers, and Sample E was grown on SiC (see supporting information for details). The circles in **Figure 1**d indicate SPEs. Isolated bright spots were found in each sample, though the density varied across the samples. We note that the samples were chosen randomly, with no specific growth requirements, to ascertain the widespread presence of SPEs in epitaxial GaN.

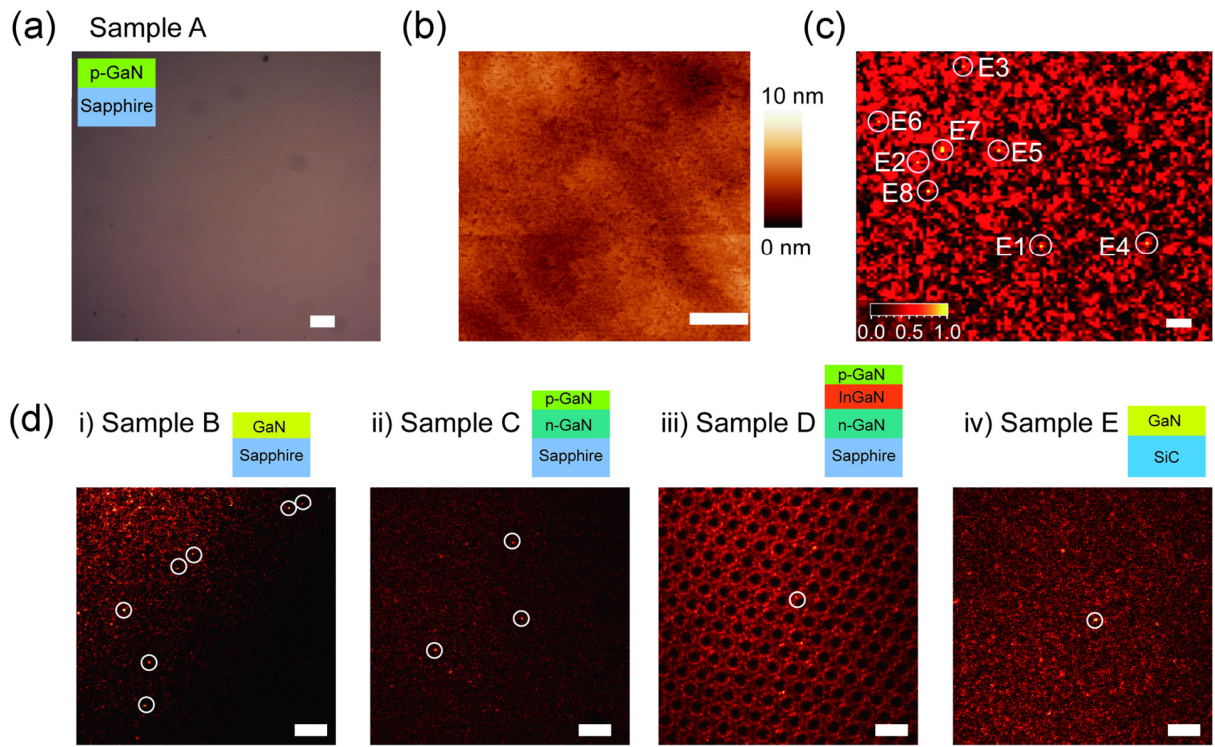

**Figure 1**. Defects in GaN wafers. a) Optical microscope image of Sample A, a 4 µm Mg- doped GaN film grown on sapphire by MOCVD. The scale bar is 10 µm. b) 10 µm x 10 µm AFM image of Sample A. The surface roughness is < 1nm. The scale bar is 2 µm. c) 60 µm x 60 µm confocal fluorescence scan of Sample A obtained using a 300 µW excitation laser. The scale bar is 5 µm. d) 40 µm x 40 µm confocal fluorescence scans of Samples B-E: i) 2 µm, undoped GaN, ii) 0.5 µm, Mg-doped GaN on 13.2 µm, Si-doped GaN, iii) GaN LED structure, iv) 1.2 µm, undoped GaN. Samples B-D are grown on sapphire. Sample E is grown on silicon carbide. The white circles indicate localized emitters. Insets: schematic diagrams of Samples A-E.

Below, we provide a detailed analysis of sample A and show representative data from the other 4 samples. As shown in **Figure 2**a, the RT photoluminescence (PL) spectra from emitters E1-E5 show distinct ZPL wavelengths of 640 nm, 657 nm, 681 nm, 703 nm and 736 nm, respectively. All spectra are obtained with the same excitation power of 100 µW, while all second-order correlation measurements were carried out using a laser power of 50 µW. The 696 nm luminescence peak corresponds to the ruby $Cr_{Al}^0$ emission and is present in all spectra taken from the samples grown on sapphire. Additional representative RT spectra from Sample A are provided in the supporting information (**Figure S2**).

The histogram in **Figure 2**b shows the ZPL wavelength distribution of 93 emitters in Sample A. This histogram bin width is 10 nm as small differences in peak position can be caused by strain variations throughout the sample[27]. However, the ZPL wavelengths span 180 nm, suggesting a different primary mechanism for the broad distribution seen in **Figure 2**b. A likely mechanism is proposed below.

The histogram in **Figure 2**c summarizes the RT ZPL line width distribution of 96 emitters in Sample A, with a median full width at half maximum (FWHM) of ~ 5 nm. Fits of the ZPL peak shape reveal an asymmetry caused by a low energy tail which may be caused by coupling to phonons. The ZPLs are narrower than the RT line width of the NV[28] center in diamond and comparable to other defects in diamond (e.g. the SiV[29] and the Cr-related[30]).

Photon emission statistics were analyzed using a Hanbury Brown and Twiss (HBT) interferometer. The spots E1-E5 in Sample A (see **Figure 1**c and **2**a) are SPEs, where the second order autocorrelation measurement $(g^2(\tau))$ at zero delay time ($\tau = 0$) shows that $g^2(0) <0.5$. Spots E6-E8 also exhibit antibunching, but the $g^2(0)$ values exceed 0.5, indicating that they are small ensembles of 2-4 emitters ($g^2(\tau)$ measurement from all 8 spots are shown in **Figure S3**).

**Figure 2**d shows spectra and autocorrelation measurements from typical emitters in Samples B-E. Data from additional emitters found in these samples is provided in the Supporting Information (**Figures S4-S7**).

To confirm that the observed fluorescence is indeed from the GaN films, three pieces of Sample A were etched using chlorine reactive ion etching (RIE) to depths of 300 nm, 4 µm and 6 µm. Subsequent PL analyses show that the emitters were still present after the 300 nm etch process, but no localized florescence was observed if the 4 µm GaN epilayer was completely removed by the etch process. This confirms that the emitters do indeed originate from GaN.

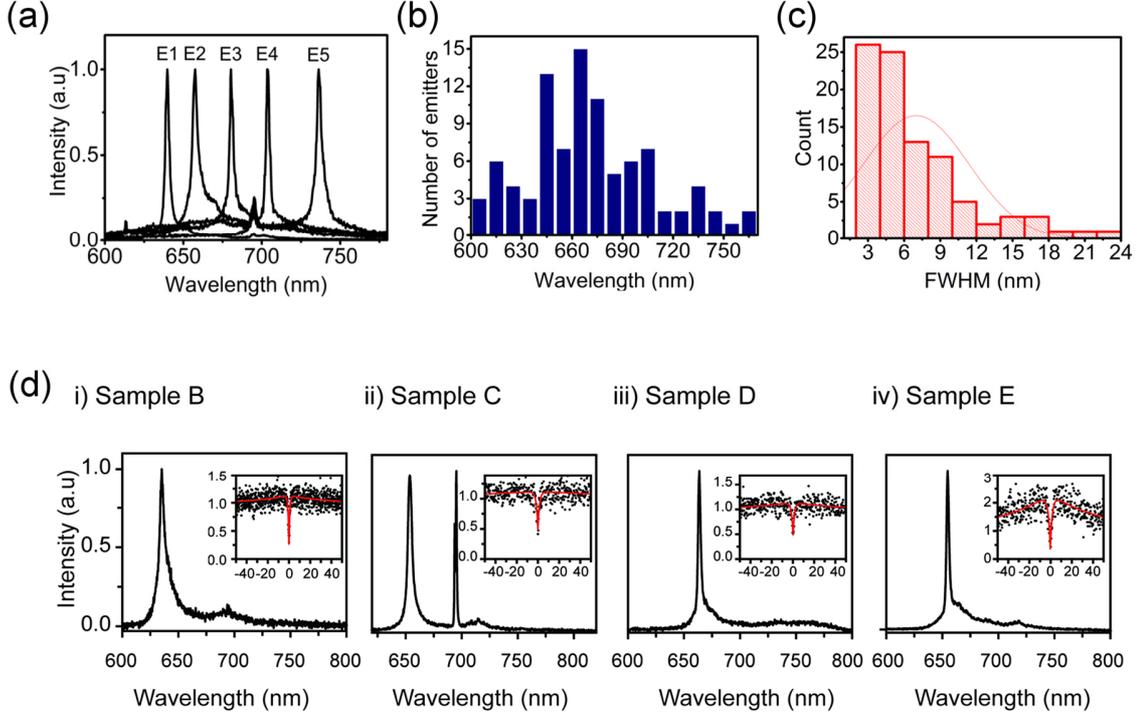

**Figure 2.** Single emitter photon emitters in GaN. a) Room temperature spectra from emitters E1-E5 (shown in **Figure 1c**) reveal distinct ZPL wavelength of 640 nm, 657 nm, 681 nm, 703 nm and 736 nm. The peak at 696 nm is the $Cr_{Al}^0$ emission from the sapphire substrate. b) Histogram of the zero phonon line wavelength distribution and c) the corresponding FWHM distribution measured from emitters in sample A. d) Typical PL spectra and $g^2$ measured from emitters in Samples B-E.

All the emitters that were investigated are bright and do not exhibit blinking or bleaching. **Figure 3**a shows the saturation behavior of emitter E2 with a ZPL wavelength of 657 nm obtained using two different methods[31]. In the first method, the saturation data are corrected for background and fitted with power model (red points and line) of the form,

$$I = I_\infty \frac{P}{P+P_{sat}}, \qquad (1)$$

where $P_{sat}$ is the saturation power and $I_\infty$ is the corresponding photon detection rate. Fitting the background-corrected data with Equation (1) yields a maximum intensity $I_\infty$ of 501×10³ counts/s at a saturation power of 930 µW. This intensity value is comparable with other single emitters in bulk materials.

Alternatively, the single photon emission rate can be estimated using $g^2$ – corrected data[31]. At an excitation power of 50 µW, Emitter E2 yields an antibunching dip at zero-delay time ($g^2_{exp}(\tau = 0)$) of 0.331. The normalized $g^2_{exp}(\tau)$ can be corrected for background using[31]:

$$g^2_{exp}(\tau) = g^2_i(\tau)\rho^2 + 1 - \rho^2 \qquad (2)$$

where $g^2_i(\tau)$ is the pure antibunching function and ρ is the ratio of single photon emission rate ($R$) to total count rate ($T$). Ideally, antibunching of a SPE such as E2 satisfies the condition $g^2_i(\tau = 0) = 0$, and the expression $R = T(1 - g^2_{exp}(\tau = 0))^{1/2}$ follows from Equation (2). Substituting experimental values for $T$ and $g^2_{exp}(\tau = 0)$ at different excitation powers, the single photon emission rate ($R$) is determined for E2. **Figure 3**a (blue dots) shows single the photon emission rate of E2 versus excitation power. Fitting this curve with the power model defined by Equation (1), a single photon emission rate of 203 x 10³ counts/s is obtained at saturation power of 313 µW. Other SPEs studied in this work yield similar values for $P_{sat}$ and $I_\infty$ (possible variations arise from excitation power dependent absorption cross-sections of different emitters). Both methods independently confirm the brightness of these emitters. A detailed analysis of the emitter photophysics and an estimate of the quantum efficiency are provided in the supporting information.

Photostability of the SPEs is studied by recording PL intensity versus time under continuous wave excitation. **Figure 3**b shows an example obtained from emitter E2 using an excitation power of 3 mW over 10 minutes of continuous acquisition. All the emitters studied in this work were stable and did not exhibit blinking or bleaching under our experimental conditions.

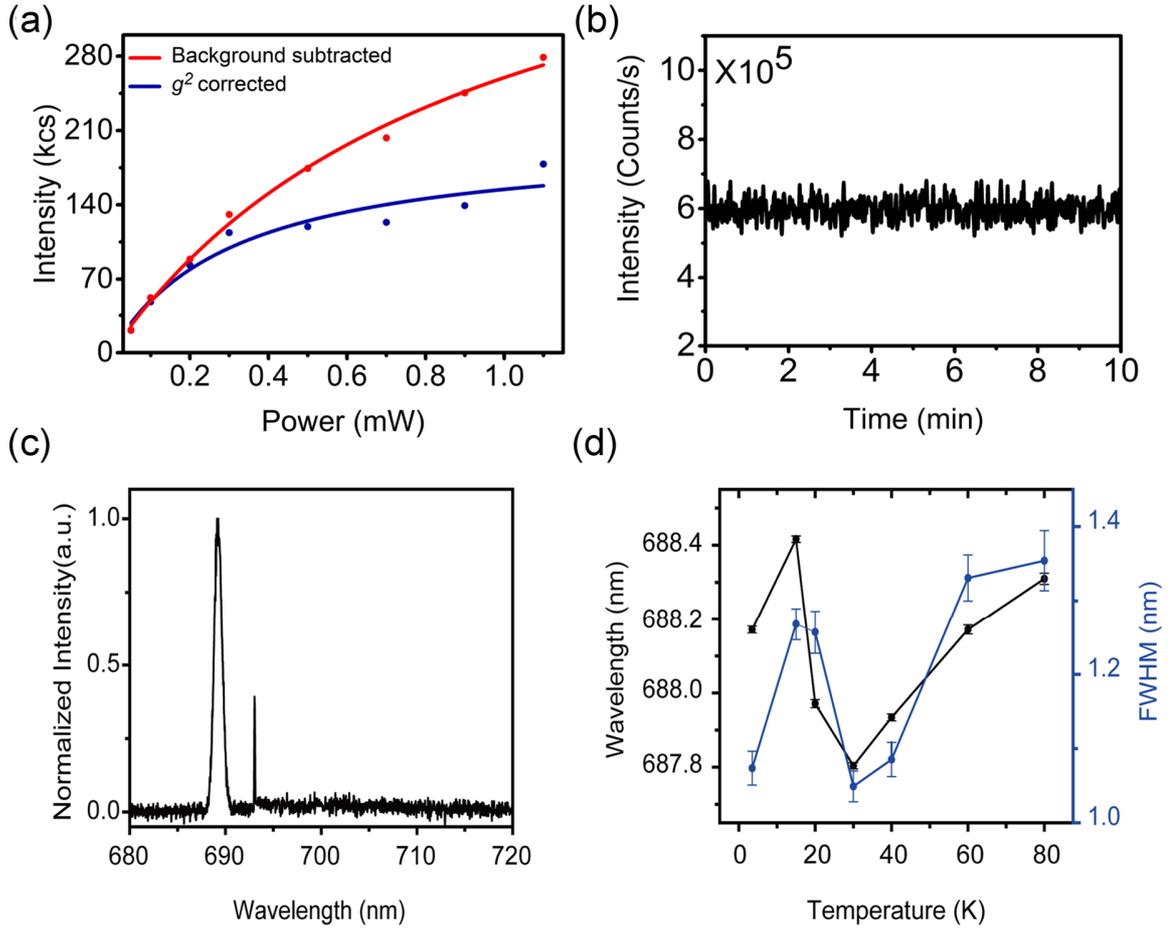

**Figure 3**. Power dependent and low temperature optical characteristics of the emitters. a) Fluorescence intensity of emitter E2 as a function of excitation power. The background-corrected saturation curve (red) yields a maximum intensity of 501 kCounts/s at a saturation power of 930 µW. The $g^2$-corrected saturation curve (blue) yields a lower bound on the maximum single photon emission rate of 203 kCounts/s at a saturation power of 313 µW. b) Long term fluorescence stability of emitter E2. The emission intensity was measured using a time bin of 50 ms. The emitter did not show any evidence of blinking or bleaching during the course of the experiments. c) PL spectrum from an emitter in Sample A acquired at 3.5 K. The peak at 693 nm is from the sapphire substrate. d) Temperature dependence of the ZPL position (black) and FWHM (blue).

We now proceed to characterize the SPEs at cryogenic temperatures. **Figure 3**c shows a PL spectrum from a SPE at ~4 K. The emitter ZPL has a FWHM of 0.98 nm corresponding to a Huang-Rhys factor of 0.46 (defined by $S=-\ln(I_{zpl}/I_{tot})$ where $I_{zpl}$ ($I_{tot}$) is the ZPL (total) integrated PL intensity[32]). We did not observe spectrometer-limited linewidths and the broadening is most likely caused by coupling to phonons or ultrafast spectral diffusion. **Figure 3**d shows the ZPL wavelength and FWHM measured as a function of temperature between 3.5 K and 80 K. Unlike other solid-state emitters that have monotonic $T^5$, $T^3$, $T^7$ or exponential dependencies on temperature[33-35], the emitters in GaN exhibit an unusual 'S-shaped' dependence, suggesting that the emitters are not simple atomic defect centers, but have an alternate origin. Furthermore, the broad spectral spread of ZPL wavelengths across individual wafers and between wafers suggests that the emitters are associated with growth defects rather than a particular impurity (such as the NV defect in diamond).

Point defect agglomeration during growth of wurtzite GaN can lead to the formation of stacking faults (SF) with varying widths spanning a few angstrom to 10 nm[36, 37]. **Figure 4**a is a schematic illustration of a 5-bilayer cubic inclusion in a 12-bilayer slab of wurtzite GaN, forming stacking faults. Consequently, localized cubic inclusions introduce a quantum well for conduction band electron due to the narrower band gap of cubic GaN relative to wurtzite GaN. The cubic inclusion is surrounded by the spontaneously polarized wurtzite matrix, resulting in electronic states with a localized electric field[38, 39]. These electronic states with a strong electric field act as effective triangular quantum well structures, altering the local optical properties of GaN[39-41]

The presence of SFs and their role in the observed quantum emissions is supported by the observed "S-shaped" temperature dependence of the ZPL wavelength seen in **Figure 3**d. The ZPL blue-shifts as T is increased from 10 K to 30 K, and red-shifts as T is increased beyond 30 K. This has been reported previously for band edge excitons in c- and a-plane grown GaN[42-44], where the blue shift is explained by an exciton transition from a shallower energy level of conduction band quantum wells to holes in the valance band. It can occur due to carrier reshuffling within the stacking faults by virtue of lattice strain or location of extrinsic atoms in the vicinity of a stacking fault[42]. The red shift at temperatures greater than 30 K can arise from thermal activation of additional bilayers, allowing deeper quantum-well-potential-bound exciton transitions[45, 46]. An alternative explanation for the red shift is a delocalization of holes in the valance band and recombination with stacking-fault-bound electrons[42, 47].

Similarly, we attribute the antibunched photon emissions reported here to the radiative recombination of an exciton bound to a point defect that resides inside or next to a stacking fault. In our model, the hole is tightly localized to the defect site whereas the electron is loosely bound by a Coulomb interaction introduced by the localized hole in the optically allowed lowest-energy excited state of the point defect. This localization modifies the stacking fault's triangular potential profile acting on the electron as illustrated in **Figure 4**b. We solve a quasi-one-dimensional Hamiltonian of this potential[48] where we applied fundamental material parameters of wurtzite and cubic GaN such as the band gaps, band alignments, effective masses of the hole and electron, dielectric constants, and steepness of the triangular well caused by spontaneous polarization. The resulting calculated energy of the exciton is assumed to be the ZPL energy of the emitter.

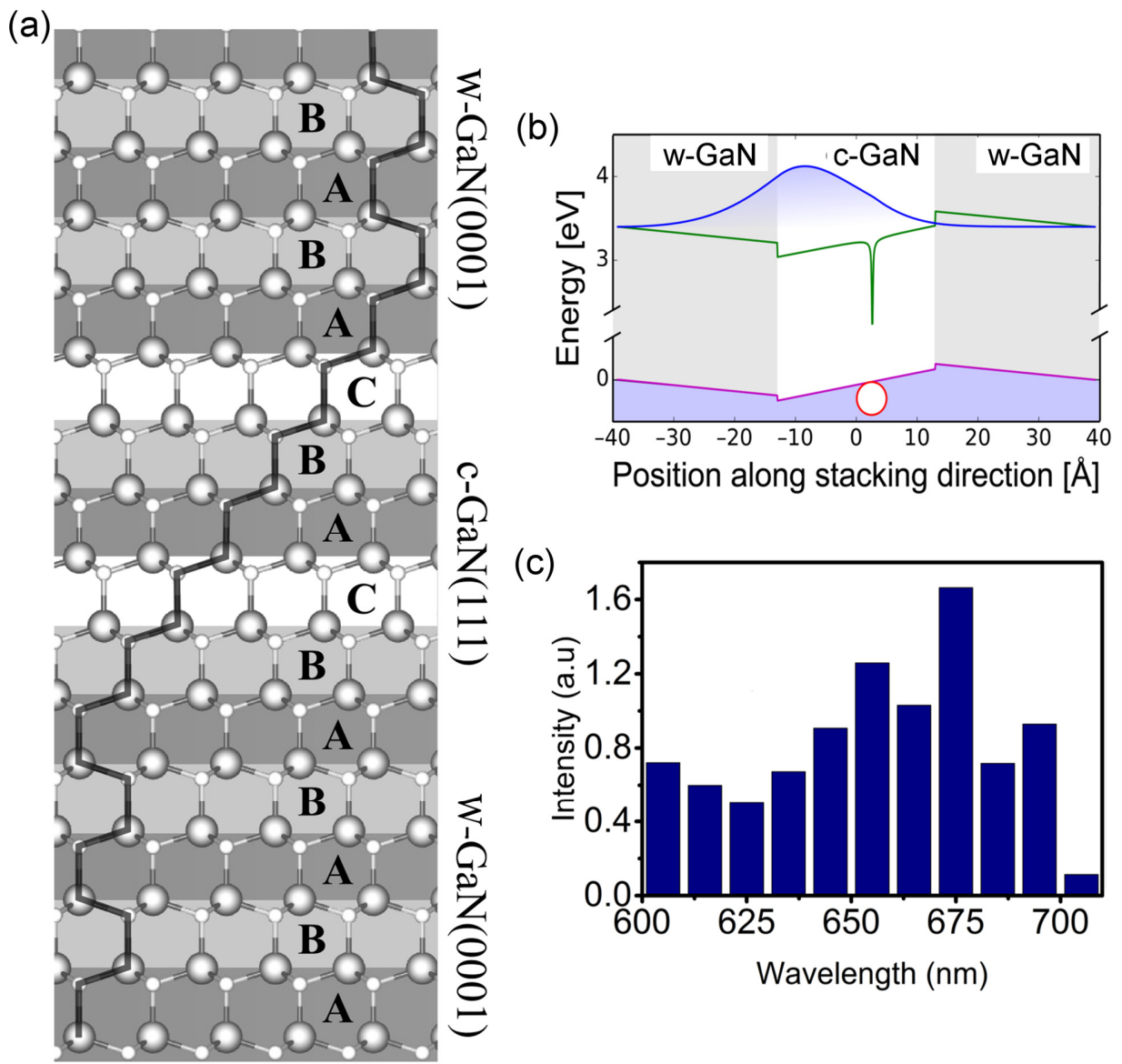

**Figure 4**. Numerical wavelength calculations: a) Schematic illustration of stacking faults generated by a cubic inclusion in wurtzite GaN. b) Location of the exciton in the cubic inclusion which spans 5 bilayers and is shown in a total slab of 12-bilayers of GaN. The potentials applied to the electron and the hole are the conduction band minimum (green curve) and valence band

maximum (purple cuve), respectively. The hole (red circle on the valence band maximum) is pinned by a point defect, while the electron is delocalized across the inclusion according to the density profile shown in blue. c) Wavelength distribution, spanning 600 to 705 nm, based on model Hamiltonian GaN parameterized calculations for a defect arrangement along the cubic inclusion shown in **Figure 4**b.

By fixing the thickness of the cubic inclusion to 5 bilayers and setting the localization potential of the hole so that the latter yields a ZPL wavelength of 680 nm for the point defect in pure wurtzite GaN, our simulation gives rise to the spectral spreading between 600 nm and 705 nm (see **Figure 4**c) if point defects are distributed uniformly between -4 nm to 4 nm with respect to the middle of the cubic inclusion. The calculated binding energy, i.e., the Coulomb-coupling in the corresponding exciton depends on the actual location of the point defects and goes up to 35 meV. The relative signal intensities are weighted according to the thermal stability of the excitons at room temperature for each defect location that lead to the final ZPL distribution of the emitters in **Figure 4**c. The modeling results are in good agreement with the experimental data, both in terms of ZPL energies and ZPL distribution. More information about the modelling is given in the supporting information.

It is important to note that the observed emitters are different from GaN or InGaN quantum dots, which have been isolated and shown to exhibit quantum emissions[49, 50]. Those sources operate predominantly at cryogenic temperatures, and originate from crystal size confinement of the QDs.

To conclude, we show that GaN is a promising host of bright SPEs in the visible and near infrared spectral ranges. In particular, we demonstrated that these emitters are prevalent in a broad range GaN films grown on sapphire and SiC substrates. Low-temperature studies and subsequent Hamiltonian parametrized calculations suggest that the quantum emitters are defects localized in close proximity to an extended stacking fault formed due to a cubic inclusion. The model suggests that generation and, potentially, control of the emitter wavelengths may be possible by intentional introduction of cubic layers in the GaN growth process. Even at the current stage, the highly photostable SPEs offer a compelling opportunity for scalable nanophotonic devices and quantum information science based on an industrially and commercially important semiconductor material.

**Acknowledgements**
The authors would like to thank Sameer Joglekar for XRD measurements and Chuck H. Choi for fluorescence-depth-scan. This work was supported in part by the U.S. Army Research Laboratory (ARL)(FA9550-14-1-0052) and the Air Force Office of Scientific Research (AFOSR) (FA9550-14-1-0052), as well as the MIT/MTL Gallium Nitride(GaN) Energy Initiative. NVT and TP would like to acknowledge the partial support of the ONR PECASE program, monitored by Dr. Paul Maki. Financial support from the Australian Research council (via DP140102721, IH150100028, DE130100592), FEI Company, and the Asian Office of Aerospace Research and Development grant FA2386-15-1-4044 are gratefully acknowledged.